\documentclass[12pt]{article}  
\usepackage{amssymb} 

\begin{document}

\title{Massive Nordstr\"{o}m Scalar (Density) Gravities from Universal Coupling}

\maketitle

\begin{center}
\author{J. Brian Pitts \\ 
John J. Reilly Center for Science, Technology, and Values,  \\ and Department of Physics \\
University of Notre Dame\\ Notre Dame, Indiana 46556 USA \\
jpitts@nd.edu, 574 904-1177 }
\end{center}

\begin{abstract}

Both particle physics  and the 1890s Seeliger-Neumann modification of Newtonian gravity suggest considering a ``mass term'' for gravity, yielding a finite range due to an exponentially decaying Yukawa potential.   Unlike Nordstr\"{o}m's ``massless'' theory, massive scalar gravities are strictly Special Relativistic, being invariant under the Poincar\'{e} group but not the  conformal group.   Geometry is a poor guide to understanding massive scalar gravities:  matter sees a conformally flat metric, but  gravity also sees the rest of the flat metric, barely, in the mass term. Infinitely many theories exhibit this bimetric `geometry,' all with the total stress-energy's trace  as  source.  All are new except the Freund-Nambu theory.   The smooth massless limit indicates  underdetermination of theories by data between massless and massive scalar gravities.    The ease of accommodating electrons, protons and other fermions  using density-weighted Ogievetsky-Polubarinov spinors in scalar gravity is noted.

\end{abstract}

keywords:  scalar gravity, Klein-Gordon equation, Nordstr\"{o}m, massive, spinor, scalar density 

\section{Introduction}

In the 1910s Nordstr\"{o}m proposed a theory of gravity that met the strictures of Special Relativity \cite{RennGenesis3,vonLaueNordstrom,OBergmannScalar}  by having, at least, Lorentz transformations as well as space- and time-translations as symmetries, as well as displaying retarded action through a field medium, as opposed to Newtonian instantaneous action at a distance. 
 Nordstr\"{o}m's scalar gravity was a serious competitor to Einstein's program for some years during the middle 1910s.  Neglecting time dependence and nonlinearity, it gives Poisson's equation just as Newton's theory does. Nordstr\"{o}m's   theory was eclipsed first by the theoretical brilliance of Einstein's much more daring project, and then by the empirical success of Einstein's theory in the bending of light, a result inconsistent with 
Nordstr\"{o}m's theory.  While representing gravity by a scalar field is no longer a viable physical proposal, it is interesting to fill a hole left  by the  abandonment of Nordstr\"{o}m's scalar gravitational theory caused by Einstein's inventing General Relativity (GR) so soon.

 Developments in group theory as applied to quantum mechanics, such as by Wigner  \cite{WignerLorentz}, 
 classified all possible fields in terms of the Lorentz group with various masses and various spins.  In the late 1930s Pauli and Fierz found that the theory of a non-interacting massless spin 2 (symmetric tensor) field in Minkowski space-time was just the linear approximation of Einstein's GR \cite{PauliFierz,FierzPauli,Wentzel}.
 Tonnelat and some author authors associated with de Broglie pursued massive spin 2 theories 
\cite{Tonnelat20}.    Nordstr\"{o}m's theory of a long-range scalar field is, in this particle physics terminology, a theory of a massless spin 0 field.

Both  the precedent of particle physics in the 1930s and the work by Seeliger and Neumann in the 1890s in giving Newtonian gravity a finite range \cite{PockelsHelmholtzEquation,Neumann,Seeliger1896,Pauli,North,NortonWoes,EarmanLambda} show the appropriateness of considering the possibility of a finite range for the gravitational potential.  A finite range corresponds in field theory to a mass term, a term in the field equation that is  linear and \emph{algebraic} in the potential; the corresponding term in the Lagrangian density, a sort of potential energy, is quadratic.

   Einstein briefly entertained in his 1917 paper on the cosmological constant what is in effect a massive scalar gravitational theory 
\begin{quote}
as a foil for what is to follow.  
In place of Poisson's equation we write
\begin{eqnarray*} 
\hspace{2cm} \nabla^2 \phi - \lambda \phi = 4 \pi \kappa \rho  \hspace{1.25cm}  .  \hspace{1.25cm}  .  \hspace{1.25cm}  .  \hspace{1.1cm} (2)
\end{eqnarray*}
where $\lambda$ denotes a universal constant.  \cite[p. 179]{EinsteinCosmological}
\end{quote}  
 Thus Einstein in effect contemplated a theory of the sort that, in light of later quantum mechanics terminology, one might call a theory of gravity using a massive scalar field, with $\lambda$ equaling the square of the scalar graviton mass in relativistic units with Planck's constant and the speed of light set to 1.  Relativistic massive scalar fields in the absence of interacting satisfy the Klein-Gordon equation, but interpreting the field as gravity introduces interactions, including self-interaction and hence nonlinearity.  %

However,  Einstein  promptly drew a widely followed  analogy to his cosmological constant---I suppress references to spare the guilty---but this analogy is erroneous \cite{DeWittDToGaF,Trautman,Treder,FMS,Schucking,CooperstockTerm,NortonWoes,HarveySchucking,EarmanLambda}.  Thus Einstein obscured for himself and others the deep conceptual issues raised by the mass term. 
The cosmological constant, having a zeroth order term in the field equations, is analogous to the scalar equation
$$\nabla^2 \phi - \lambda (1+\phi) = 4 \pi \kappa \rho;$$
the strange  term $- \lambda \cdot 1$ will tend to dominate over the intended  $- \lambda \phi$ term.

 Two papers from \emph{c.} 1970 provide a partial exception  to the remarkable silence about the possibility of relativistic massive scalar gravity theories.  One by Freund and Nambu \cite{FreundNambu} writes down equations that formally could be read as applying to massive scalar gravity, but they do not consider that application.   Deser and Halpern  \cite{DeserHalpern} soon called attention to the identity of the Freund-Nambu field equations (less the mass term, though that omission is not mentioned!) with Nordstr\"{o}m's  theory.  While the alert reader could notice that Freund and Nambu had in effect provided field equations for a massive variant of Nordstr\"{o}m's theory, apparently no one has ever managed to comment on  that fact, still less to discuss its significance.  A more recent paper by H. Dehnen and R. Frommert reinvented massive scalar gravity in a similar way \cite{DehnenMassiveScalar}, but with an unfortunate and apparently unnoticed restriction on the allowed matter field content such that standard scalar fields are inadmissible.

Massive scalar gravities, if the mass is sufficiently small, fit the data as well as does Nordstr\"{o}m's theory, as a consequence of the smoothness of the limit of a massive scalar field theory as the mass goes to zero  \cite[p. 246]{WeinbergQFT1}.  
 Thus there is a problem  of underdetermination between the massless theory and its massive variants for sufficiently small masses \cite{UnderdeterminationPhoton}.  The analog of this instance of underdetermination was already clearly understood by Seeliger in the 1890s. He wrote  (as translated by John Norton) that Newton's law was ``a purely empirical formula and assuming its exactness would be a new hypothesis supported by nothing.'' \cite{Seeliger1895a,NortonWoes}
While that claim might be a bit strong, in that Newton's law had virtues that not every rival formula empirically viable in the 1890s had, a certain kind of exponentially decaying formula of Neumann and Seeliger was also associated with an appropriate differential equation \cite{PockelsHelmholtzEquation,Neumann}.

It is well known that Nordstr\"{o}m's theory does not bend light \cite{Kraichnan}.  That is an immediate consequence of the conformal flatness of the metric in Nordstr\"{o}m's theory  in geometrical form \cite{EinsteinFokker,EinsteinTrans4} 
 and the conformal invariance of Maxwell's electromagnetism \cite{Wald}: space-time is flat in Nordstr\"{o}m's theory except for the volume element, but light doesn't see the volume element in Maxwell's theory.  
 While scalar gravity is a museum piece as far as theoretical physics is concerned---at least as far as the dominant gravitational field as concerned---it remains a useful test bed theory for analogous phenomena for which the details in General Relativity are much more complicated technically  or might play a secondary role in gravitation theory 
\cite{BransScalar,MisnerScalar,SundrumScalar,ReuterManrique}.  Scalar gravity  sheds  light on some fundamental issues in space-time theory as well, not least by setting a precedent that could be entertained for massive tensor theories.  




\section{Massive Scalar Gravities:  Relatives of Nordstr\"{o}m's Theory }

  Here I shall give a suitable Lagrangian density in a form adapted to the derivation of the massive variants to be introduced shortly.
The Einstein-Fokker geometrization  \cite{EinsteinFokker,EinsteinTrans4} 
 suggests a useful set of variables to use.  One can isolate the conformal structure (the null cones) out of a metric by taking the part with determinant of $-1$; for a flat metric $\eta_{\mu\nu}$ one can call the resulting conformal metric density $\hat{\eta}_{\mu\nu},$  a tensor density of weight $-\frac{1}{2}$ in four space-time dimensions \cite{Anderson}. 
Let $\tilde{\eta}$ be a (positive) scalar density of arbitrary nonzero weight $w,$ related to $\eta_{\mu\nu}$ by 
$\sqrt{-det(\eta_{\mu\nu})} = \tilde{\eta}^\frac{1}{w}.$      
(The expression $\sqrt{-det(\eta_{\mu\nu})}$ is often written as $\sqrt{-\eta}.$) 
Thus $\tilde{\eta}$ (or its $w$th root) governs volumes, at least volumes that are not distorted by gravity.  Let the gravitational potential be represented by a potential $\tilde{\gamma},$ also of density weight $w.$ Then one can define a new effective volume element by 
$ \tilde{g} =_{def} \tilde{\eta} + k \tilde{\gamma}.$ Thus far it is unclear whether $ \tilde{g}$ actually determines the volume of anything, but the derivation shows that it determines all volumes for the matter dynamics and   all (in the massless case) or most (in the massive cases) for the gravitational dynamics.  
It turns out that $k^2 = 64 \pi G w^2;$ the sign of $k$ does not matter much, but will be chosen to match that of $w$ to maximize continuity.  
Neglecting terms that do not contribute to the field equations or that disappear with the choice of appropriate coordinates, one can take the purely gravitational part of the Lagrangian density  as 
\begin{eqnarray}
\mathcal{L}_{g0}= - \frac{1}{2} \hat{\eta}^{\mu\nu} \tilde{g}^{ \frac{1}{2w} -2} (\partial_{\mu} \tilde{\gamma} ) \partial_{\nu} \tilde{\gamma}; 
\end{eqnarray}  
here $\partial_{\mu}$ is the coordinate derivative, while $\hat{\eta}^{\mu\nu}$ is the inverse of $\hat{\eta}_{\mu\nu}.$ 
This result will be derived below. 
It is essential to use scalar \emph{densities} in order to obtain the trace of the stress-energy tensor so readily, which then permits  combining the gravitational potential with the background volume element by an additive field redefinition $ \tilde{g} =_{def} \tilde{\eta} + k \tilde{\gamma}.$ 
The basic postulate is universal coupling, that the full field equations are obtained by taking the free field equations and adding in the trace of the total stress-energy tensor (including gravitational energy-momentum).
The universal coupling principle (initially using a brute-force direct construction of the stress-energy tensor)
was employed by Einstein in his supposedly unsuccessful \emph{Entwurf} physical strategy for finding his field equations  \cite{EinsteinEntwurf}, and was later brought to successful completion in finding Einstein's equations using higher mathematical technology \cite{Kraichnan}.
One can show that the trace of the stress-energy tensor is given by taking the Euler-Lagrange derivative of the Lagrangian density  
with respect to the volume element (perhaps raised to some power), such as
\begin{equation}
\frac{\delta \mathcal{L}}{\delta \sqrt{-\eta} } = \frac{1}{2\sqrt{-\eta}} \frac{\delta \mathcal{L}}{\delta \eta_{\mu\nu} }   \eta_{\mu\nu}.
\end{equation}
The universal coupling postulate can be written therefore as  
\begin{eqnarray} \frac{\delta \mathcal{L} }{\delta \tilde{\gamma} } = \frac{\delta \mathcal{L}_{free} }{\delta \tilde{\gamma} } 
+ k \frac{\delta \mathcal{L}}{\delta \tilde{\eta} }|\tilde{\gamma}. \end{eqnarray}
The same theory results from all values of $w$.   $\mathcal{L}_{free}$ is a rather standard quadratic expression yielding the Klein-Gordon equation.  The material part of the Lagrangian density can be written in terms of matter fields $u$ and the conformally flat metric built from  $\hat{\eta}_{\mu\nu}$ and $\tilde{g}$; $\tilde{\eta}$ does \emph{not} appear on its own.  Both the gravitational and the material parts of the Lagrangian density therefore depend only on a conformally flat metric.

While Nordstr\"{o}m's theory has been derived in terms of universal coupling previously \cite{Kraichnan,FreundNambu,DeserHalpern}, the approach outlined here admits a ready generalization yielding a distinct massive scalar gravity for every 
 value of $w,$ only one of which has been found before.   The kinetic and matter terms are of course just those of Nordstr\"{o}m's theory.  Given the nonlinearity of the theories, there are different choices of field variables, often nonlinearly related, that are especially convenient for one purpose or another; thus comparison requires choosing some common set of fields.  
It is convenient to use 
$\sqrt{-\eta}= \tilde{\eta}^\frac{1}{w}$ and $\sqrt{-g}= \tilde{g}^\frac{1}{w}$; here $\eta$ and $g$ (without the $~\tilde{\ }$) are the determinants of the metrics $\eta_{\mu\nu}$ and $g_{\mu\nu}$ as usual.     In the mass term only, one  has  $\sqrt{-\eta}$ appearing in its own right, the fact around which much of the philosophical interest of the theories.  For any  real $w$ (including $w=1$ and $w=0$ by  l'H\^{o}pital's rule), a universally coupled massive variant of Nordstr\"{o}m's theory is given by 
\begin{eqnarray}
\mathcal{L}_{mass} =  \frac{m^2}{64 \pi G} \left[ \frac{  \sqrt{-g} }{w-1}   +   \frac{ \sqrt{-g}^w \sqrt{-\eta}^{1-w} }{w(1-w)}  -  \frac{ \sqrt{-\eta}  }{w} \right].
\end{eqnarray}
One can express this mass term as a quadratic term in the potential  and, typically, a series of higher powers using the expansion $$ \tilde{g} =\sqrt{-g}^w = \sqrt{-\eta}^w + 8 w \sqrt{\pi G} \tilde{\gamma},$$ where $\tilde{\gamma}$ is the gravitational potential; the case $w=0$ requires taking the $\frac{1}{w}$th root of this equation and taking the limit, giving an exponential function.  In all cases the result is
\begin{eqnarray}
\mathcal{L}_{mass} = 
 -m^2   \left[ \frac{ \tilde{\gamma}^2 }{ 2 \sqrt{-\eta}^{ 2w-1} }   +  \frac{ (1-2w) 4 \sqrt{\pi G} \tilde{\gamma}^3 }{3 \sqrt{-\eta}^{3w-1}  }  + \ldots \right].
\end{eqnarray}
This one-parameter family of theories closely resembles the 2-parameter family Ogievetsky-Polubarinov family of massive tensor theories \cite{OP,OPMassive2}, which can also be derived in a similar fashion \cite{MassiveGravity1}. The case $w= \frac{1}{2},$ which conveniently terminates at quadratic order, is the Freund-Nambu theory \cite{FreundNambu}. It is also useful to make a series expansion for all the theories using the special $w=0$ exponential variables, as will appear below.


\section{Massive Scalar Gravities Are Strictly Special Relativistic}

Features of Nordstr\"{o}m's scalar gravity  are  said to have shown that even the simplest and most conservative relativistic field theory of gravitation burst the bounds of Special   Relativity (SR)   \cite[p. 179]{MTW}  \cite[p. 414]{NortonNordstromBook}.  
Relativistic gravity couldn't be merely special relativistic, according to this claim.  
Nordstr\"{o}m's theory indeed has a merely conformally flat space-time geometry \cite{EinsteinFokker},
which one can write as 
\begin{equation} g_{\mu\nu}=\hat{\eta}_{\mu\nu} \sqrt{-g}^\frac{1}{2}, \end{equation} 
where $\hat{\eta}_{\mu\nu}$ (with determinant $-1$) determines the light cones just as if for a flat metric in SR.  Thus 
Nordstr\"{o}m's theory is invariant under the 15-parameter conformal group, a larger group than the usual 10-parameter Poincar\'{e} group of Special Relativity.   

 By contrast the massive variants of Nordstr\"{o}m's theory are invariant only under the 10-parameter Poincar\'{e} group standard in SR and thus are special relativistic in the strict sense. The mass term breaks the conformal symmetry. It is therefore false that relativistic gravitation could not have fit within the confines of Special Relativity.  While it is true that no phenomena required the mass term, it was  possible that the mere smallness of the mass parameter explained its empirical obscurity, as Seeliger had already proposed in the Newtonian case.

\section{Derivation of (Massless) Nordstr\"{o}m   Scalar Gravity }

Having summarized the results above, I turn to their derivation.
While Nordstr\"{o}m's theory has been derived in terms of universal coupling previously \cite{Kraichnan,FreundNambu,DeserHalpern}, the approach outlined here admits a ready generalization yielding a distinct massive scalar gravity for every value of $w.$ All of these theories are new, with the exception of the Freund-Nambu theory.  The use of a scalar \emph{density} as the field variable permits a convenient additive change of variables and simple coupling to the trace of the total stress-energy tensor.  A scalar density under arbitrary coordinate transformations is of course a scalar under Lorentz transformations (with two options regarding transformations with negative determinant \cite{Golab}, corresponding to the scalar and pseudoscalar of particle physics).  It will turn out that different density weights conveniently give different massive scalar gravities, much as in the tensor case \cite{OP,MassiveGravity1}.  

\subsection{Tensor Densities and Irreducible Parts of the Metric and Stress Tensor }
%
   In addition to the background geometrical variables and the gravitational potential, there are also other matter fields, which I denote collectively by $u.$  
The derivation that follows makes use of the Rosenfeld-type  metrical definition of the stress-energy tensor \cite{RosenfeldStress,Kraichnan,Deser,GotayMarsden,SliBimGRG}. In such an approach the flatness of the metric is momentarily relaxed while a functional derivative with respect to it is taken; then flatness is restored.  It will be most helpful to use not the metric tensor $\eta_{\mu\nu},$ its inverse, or some densitized relative thereof, but rather two irreducible geometrical objects that together build up the metric tensor. (The use of irreducible geometric objects is also important in analyzing the Anderson-Friedman geometric objects program \cite{FriedmanJones}.) The conformal metric tensor density $\hat{\eta}_{\mu\nu}$ (with determinant of $-1$) of weight $-\frac{1}{2}$ determines the null cone structure, which is untouched by gravitation in scalar gravity theories.  The remainder of the flat metric tensor is supplied by a scalar density $\tilde{\eta}$ of nonzero weight $w$, which quantity we may take to be positive.\footnote{Choosing  $\tilde{\eta}>0$ in all coordinate systems indicates which of the various subtypes of density \cite{Golab} is intended.} The flat metric tensor is built up as $\eta_{\mu\nu}= \hat{\eta}_{\mu\nu} (\tilde{\eta})^\frac{1}{2w}.$  Note that $w$ can be any real nonzero number; the use of a single tilde above the variable name reminds us that the field is a density, but gives no hint about which weight it has.  The usual torsion-free covariant derivative compatible with $\eta_{\mu\nu}$  is written as $\partial_{\alpha}.$  The metric-compatibility condition $\partial_{\alpha}  \eta_{\mu\nu}=0 $ yields $\partial_{\alpha} \tilde{\eta}=0$ and $\partial_{\alpha} \hat{\eta}_{\mu\nu}=0$ as well.  One should be careful to use the correct form of the covariant derivative for tensor densities, so it is worth recalling the forms of their covariant and Lie derivatives.  A $(1, 1)$ density $\tilde{ \phi }^{\alpha}_{\beta}$ of weight $w$ is representative.  The Lie derivative is given by \cite{Schouten,Anderson,Israel}
\begin{eqnarray}
\pounds_{\xi} \tilde{ \phi }^{\alpha}_{\beta} = \xi^{\mu} \tilde{ \phi } ^{\alpha}_{\beta},_{\mu} -
\tilde{ \phi }^{\mu}_{\beta} \xi^{\alpha},_{\mu} + \tilde{ \phi }^{\alpha}_{\mu} \xi^{\mu},_{\beta} + w
\tilde{ \phi }^{\alpha}_{\beta} \xi^{\mu},_{\mu},
\end{eqnarray}  where the $,\mu$  denotes partial differentiation with respect to local coordinates $x^{\mu}.$ 
The $\eta$-covariant derivative is given by  \cite{Schouten,Anderson,Israel} 
\begin{eqnarray}   
\partial_{\mu} \tilde{ \phi }^{\alpha}_{\beta} = \tilde{ \phi }^{\alpha}_{\beta},_{\mu} + \tilde{ \phi }^{\sigma}_{\beta}
\Gamma_{\sigma\mu}^{\alpha} - \tilde{ \phi } ^{\alpha}_{\sigma} \Gamma_{\beta\mu}^{\sigma} - w
\tilde{ \phi }^{\alpha}_{\beta} \Gamma_{\sigma\mu}^{\sigma}.
\end{eqnarray} 
Here $\Gamma_{\beta\mu}^{\sigma}$ are the Christoffel symbols for $\eta_{\mu\nu}.$  Once the curved metric $g_{\mu\nu}$ is defined, the analogous $g$-covariant derivative $\nabla$ with Christoffel symbols $\{ _{\sigma\mu}^{\alpha} \}$ follows.   For \emph{scalar} densities, the relevant piece is the new term with the coefficient $\pm w.$  
The formulas for Lie and covariant differentiation follow \cite{SzybiakLie,SzybiakCovariant}  from the coordinate transformation law for scalar densities: under a change of local coordinates from an unprimed 
 set $x^{\mu}$ to a primed set $x^{\nu^{\prime} },$ the density's component behaves as 
\begin{equation} 
\phi^{\prime} = \left| det\left( \frac{ \partial x }{\partial x^{\prime} }\right) \right|^{w} \phi,
\end{equation}
The primes perhaps are opposite where one might have expected, but this is the usual convention  \cite{Anderson,Golab,Schouten}, though some authors, especially but not only Russian, 
define weight in the opposite fashion.  For the other kind of scalar densities, which can change sign under coordinate transformations \cite{Golab}, the same Lie and covariant derivative formulas follow, because only behavior for infinitesimal transformations is relevant.

For  any action  $S$ invariant under arbitrary coordinate transformations, one can derive a metrical stress-energy tensor. It is convenient to break the flat metric up into its irreducible parts, the conformal metric density $\hat{\eta}_{\mu\nu}$ that fixes the null cones and the volume-related scalar density $\tilde{\eta}$ \cite{SliBimGRG}. One can show that 
 $$ \frac{\delta S}{\delta \tilde{\eta} } = \frac{1}{2w \tilde{\eta} } \frac{ \delta S}{\delta \eta_{\mu\nu} } \eta_{\mu\nu}.$$
 Making an infinitesimal coordinate transformation described by the vector field $\xi^{\mu}$ gives
$$ \delta S = \int d^{4}x \left[ \frac{ \delta S}{\delta \tilde{\gamma} } \pounds_{\xi} \tilde{\gamma} +  
\frac{ \delta S}{\delta u } \pounds_{\xi} u +   
\frac{ \delta S}{\delta  \hat{\eta}_{\mu\nu} } \pounds_{\xi} \hat{\eta}_{\mu\nu} +
\frac{ \delta S}{\delta  \tilde{\eta} } \pounds_{\xi} \tilde{\eta} \right]  + BT,$$ where BT is some boundary term of no interest for present purposes.  Because $S$ is a scalar, $\delta S = 0.$  Integration by parts pulls all the derivatives off $\xi^{\mu}$ at the cost of more boundary terms.  Choosing  $\xi^{\mu}$ to vanish at the boundary annihilates all the boundary terms, while using its arbitrariness removes the integration, leaving the integrand to vanish.  Going `on-shell' by using gravity's and matter's field equations $\frac{ \delta S}{\delta \tilde{\gamma} } =0,$ $\frac{ \delta S}{\delta u } =0,$ respectively, gives local conservation of stress-energy:
$$-\partial_{\nu} \left[2 \hat{\eta}_{\mu\alpha} \frac{ \delta S}{\delta  \hat{\eta}_{\mu\nu} }
+ w \delta^{\nu}_{\alpha} \tilde{\eta} \frac{ \delta S}{\delta  \tilde{\eta} }\right] =0.$$
The stress-energy tensor here is broken up into traceless and trace parts.  For a scalar theory, the latter will represent the source for gravity.  This stress-energy tensor contains contributions from both matter $u$ and gravity $\tilde{\gamma} $. If Maxwell's electromagnetism is included among the matter fields, then it couples only to $\hat{\eta}_{\mu\alpha},$ not $\tilde{\eta}$:  rather than introducing conformal rescalings and showing that nothing interesting changes \cite{Wald,ChoquetDeWitt2}, this use of densities is the most direct way to see the theory's conformal invariance, from which the theory's failure to bend light follows immediately:  electromagnetic radiation in the absence of charges doesn't know that it isn't in flat space-time because the volume element disappears entirely. 


\subsection{Spinors}

Foundational questions about space-time not infrequently are addressed  as though there were no such thing as protons, electrons, neutrons, or other fermions, which are represented classically by spinor fields, or as if spinor fields introduced no additional issues.  But consider the supposedly \emph{trivial} possibility of representing special relativity using arbitrary coordinates.  While mere tensor calculus is needed for bosonic fields, the issue is more complicated for spinor fields.  In the interest of overcoming this unjustifiable neglect of spinor fields, I point out that 
spinorial matter can be included effortlessly in the universally coupled massive gravities considered here.  
On account of breaking up the metric and stress tensor into their irreducible pieces, including spinors requires no work at all, because the spinor (if massless) does not notice scalar gravity.  To see this effortlessly and yet rigorously, one can use the Ogievetsky-Polubarinov spinor formalism \cite{OP,OPspinor,GatesGrisaruRocekSiegel,BilyalovSpinors}, thereby avoiding a tetrad in favor of the metric (which is possible by construction, contrary to widely held belief); the formalism is a bit like the tetrad formalism in the symmetric gauge, but is conceptually independent. The conformal covariance of the massless Dirac equation \cite{ChoquetDeWitt2,Branson}  is well known.  But what is rarely if ever noticed is that 
one must and can use density-weighted spinors to achieve conformal \emph{invariance}, with the volume element dropping out altogether---much as in Maxwell's electromagnetism, but the details are more difficult and less familiar and involve derivatives of the conformal part of the metric \cite{PittsPhilDiss}.  The trace of the stress-energy tensor comes from $\frac{\delta S}{\delta \tilde{\eta} },$ but $\tilde{\eta}$ is simply absent from the suitably weighted spinor's kinetic term.  That expression depends, in a highly nonlinear way, only on the components of $\hat{\eta}_{\mu\nu},$ which determines the null cones:  9 components, not the 10 of the metric, 15 of a unimodular conformal tetrad, or 16 of a tetrad.   The appropriate spinor has density weight $\frac{3}{8}$ in 4 space-time dimensions  or $ \frac{n-1}{2n}$ in $n$ space-time dimensions \cite{PittsPhilDiss}.  The spinor, if massless, does not notice scalar gravity at all; the metrical stress-energy tensor has vanishing trace even off-shell.   More familiar routes to this conclusion of vanishing trace of the metric stress tensor are  less direct \cite{ImprovedEnergy,SorkinStress,DehnenHiggsScalar}.
 The Ogievetsky-Polubarinov treatment of spinors has also avoided a spurious counterexample to the Anderson-Friedman absolute objects program \cite{FriedmanJones} for understanding the difference between merely formal general covariance and the substantive kind that is supposed to be a novel feature of General Relativity.

\subsection{Universal Coupling}

Let us assume that the free field action $S_{f}[ \tilde{\gamma}, u, \hat{\eta}_{\mu\nu}, \tilde{\eta}]$ (with vanishing Newton's constant $G$) is known; it is given below, apart from the unspecified matter fields. The full action $S$ should reduce to $S_{f}$ in the limit of vanishing $G.$   The task at hand is to derive the full action $S$ for the theory with nonzero gravitational interaction.  Lorentz covariance requires a source generalizing the mass density in a Lorentz-invariant way, so the trace of the stress-energy tensor is a natural choice.  This choice is perhaps not compulsory, unlike the tensor case where free field gauge invariance necessitates that any source used by a divergenceless symmetric rank 2 tensor, of which there is only one physically significant example at hand.  But it is a very natural choice.

Letting $S$ be specialized once more to the interacting scalar gravitational theory that we seek, we can postulate that the gravitational free field equation is modified by the introduction of a source term that is basically the trace of the stress-energy tensor:
$$ \frac{ \delta S}{\delta \tilde{\gamma} } = \frac{ \delta S_f}{\delta \tilde{\gamma} } + k 
\frac{ \delta S}{\delta  \tilde{\eta} }|\tilde{\gamma} ,$$ where $k$ is a coupling constant related to $G$ and perhaps the density weight $w$ in ways that will be ascertained later. In anticipation of a change to bimetric variables, the $|\tilde{\gamma}$ notation has been added to emphasize that the other independent variable here besides $ \tilde{\eta}$ is $\tilde{\gamma}$. 

The change to bimetric variables\footnote{ For the tensor case, the new variables involve two metric tensors or the like \cite{Kraichnan,SliBimGRG,MassiveGravity1}.  For the scalar case, only the scalar density portions undergo any redefinition, because the gravitational field has no tensor piece to combine with $\hat{\eta}_{\mu\nu}.$  Nonetheless the term ``bimetric variables'' is handy.}
 involves the definition
$$\tilde{g} = \tilde{\eta} + k \tilde{\gamma}.$$  Equating coefficients for variations of $\tilde{\eta}$ and $\tilde{\gamma}$ relates the functional derivatives as follows:
$$   \frac{ \delta S}{\delta \tilde{\eta} }|\tilde{\gamma}=  \frac{ \delta S}{\delta \tilde{g} } + \frac{ \delta S}{\delta \tilde{\eta} }|\tilde{g}; $$
$$\frac{ \delta S}{\delta \tilde{\gamma} }= k \frac{ \delta S}{\delta \tilde{g} }. $$
The second equation shows that the new field $\tilde{g}$ has an Euler-Lagrange equation equivalent to that of the potential $\tilde{\gamma}.$  The first equation shows that the trace of the stress-energy tensor splits into one piece that vanishes on-shell and one that does not.  Using these equations in the postulated equation for universal coupling gives
$$ 0= \frac{ \delta S_f}{\delta \tilde{\gamma} }  + k 
\frac{ \delta S}{\delta  \tilde{\eta} }|\tilde{g}.$$

So far little has been said about the detailed form of $S_f$.  The most natural choice is given by the Lagrangian density 
$$\mathcal{L} = -\frac{1}{2} (\partial_{\mu}\tilde{\gamma}) (\partial_{\nu}\tilde{\gamma}) \hat{\eta}^{\mu\nu} \tilde{\eta}^{\frac{1}{2w} -2},$$ apart from a mass term for $\tilde{\gamma}$ which will be included later (and a free field term for matter $u$ which does not contain $\tilde{\gamma}$ and so does not contribute to the derivation).  This choice gives the usual wave equation.  The metrical signature $-+++$ is employed.

To satisfy the universal coupling identity in bimetric guise, it is convenient \cite{Kraichnan,Anderson,SliBimGRG}  to split the full (but unknown) action $S$ into a piece $S_{1}[\tilde{g}, u, \hat{\eta}_{\mu\nu}] $ (without explicit dependence on $\tilde{\eta}$) and another piece $S_2$ that, perhaps among other things, cancels the term $\frac{ \delta S_f}{\delta \tilde{\gamma} }$.  
For $S_1$ it is natural to build a conformally flat metric $$g_{\mu\nu}=  \hat{\eta}_{\mu\nu} (\tilde{g})^\frac{1}{2w}.$$  Then it is natural to choose the Hilbert-like expression $$S_{1}= c \int d^{4}x \sqrt{g}R[g] + S_{matter}[g_{\mu\nu},u].$$ (A cosmological constant term $\int d^4 x \sqrt{-g}$ is also available if desired.)  
One can show that $$S_{2}=\frac{2w}{3k} \int d^{4}x R[\eta] \tilde{\eta}^{-1 + \frac{1}{w}} \tilde{\gamma}  + \int d^{4}x \partial_{\mu} \alpha^{\mu}$$ does the job of accommodating $S_{f}.$  The first piece $S_1$involving the Ricci scalar for $\eta_{\mu\nu}$ does the work here.  The second piece $S_{2}$ is simply a boundary term, which is deposited into $S_2$ rather than elsewhere for convenience. 
The boundary term can be chosen to remove the second derivatives from the Hilbert-like term $\sqrt{g}R[g]$ in $S_1.$  One can also include a pure volume term  $\int d^4 x \sqrt{-\eta}$.   Note that $S_{2}$ contributes nothing to the field equations; its purpose is to contribute to the Rosenfeld metric stress-energy tensor only. (Recall that flatness of the background is relaxed briefly in taking the functional derivative and then restored.)    The total action for the massless case is thus a piece $S_1$ describing an effective conformally flat geometry and a piece $S_{2}$ that does not affect the field equations.  Universal coupling has completely clothed the background volume element  with the gravitational potential, leaving only their sum as observable.   This is an amusingly strict realization of Einstein's Poincar\'{e}-inspired dictum that only the epistemological sum of gravity and physics is observable \cite{EinsteinGeomExp}.  
By requiring the usual normalization for the free gravitational field's kinetic term to lowest order, one infers that $$ c= - \frac{4w^2}{3k^2}.$$
In comparison to the free gravitational Lagrangian density  
\begin{eqnarray}
 - \frac{1}{2} \hat{\eta}^{\mu\nu} \tilde{\eta}^{ \frac{1}{2w} -2} (\partial_{\mu} \tilde{\gamma} ) \partial_{\nu} \tilde{\gamma}, 
\end{eqnarray}  
the interacting theory has the corresponding expression (apart from terms not affecting the equations of motion)
\begin{eqnarray}
 - \frac{1}{2} \hat{\eta}^{\mu\nu} \tilde{g}^{ \frac{1}{2w} -2} (\partial_{\mu} \tilde{\gamma} ) \partial_{\nu} \tilde{\gamma}, 
\end{eqnarray}  
while the matter fields see the conformally flat effective metric rather than the flat background metric.  Thus universal coupling of gravity to the trace of the total stress-energy tensor yield's Nordstr\"{o}m's theory.  In the massless case, the same theory obtains for every (nonzero) density weight $w$   of the gravitational potential.   The case $w=0$ without a graviton mass term was handled by Kraichnan \cite{Kraichnan} using an multiplicative exponential field redefinition, rather than an additive one, and yields Nordstr\"{o}m's theory as well.

\section{Derivation of Massive Gravities}


Making the gravitational potential $\tilde{\gamma}$ massive implies that the free gravitational potential obeys the Klein-Gordon equation, 
 not the wave equation (or Laplace's equation in the static case).  It is therefore necessary to add a term quadratic in the gravitational potential to the free gravitational field Lagrangian density.  Going through the universal coupling derivation, one then fields a crucial new term in which the flat metric's volume element remains essentially in the theory.  One also crucially uses the cosmological constant term; the cosmological constant term and the new term with the flat volume element have equal and opposite terms linear in the gravitational potential in the total action, a crucial cancellation that removes the odd behavior (from a field-theoretic point of view \cite{FMS}) of the cosmological constant.  Then the pure volume term enters the action to cancel out the zeroth order parts of both the cosmological constant and the new term essentially involving the flat volume element; this last cancellation is largely  a matter of good bookkeeping, letting the action vanish for flat space-time but not affecting the field equations.  Thus the interacting theory has a term quadratic in the gravitational potential (a mass term) and, except in one special case, one or more higher powers (possibly infinitely many) in the gravitational potential.  While the derivation is carried out using different fields for different values of density weight $w,$ making comparison mildly nontrivial, one can show (such as by using the two metrics only) that the theories obtained are all distinct.  One can also show that the resulting one-parameter family of mass terms is related in the expected way to the two-parameter family of massive tensor gravities obtained some time ago by Ogievetsky and Polubarinov.  (It turns out that the mass term in \cite{OP} contains a typographical error absent in the less well known summary \cite{OPMassive2}.)

One expects that the mass term for a free field be quadratic in the potential and lack derivatives.  The free field action $S_{f}= S_{f0} + S_{fm}$ is now assumed to have two parts: a (mostly kinetic) part $S_{f0}$ that as in the massless case above, and an algebraic mass term $S_{fm}$ that is quadratic.  We seek a full universally coupled theory with an action $S$ that has two corresponding parts.  The two parts of $S=S_{0} + S_{ms}$ are the familiar part $S_{0}$ (yielding the Einstein tensor, the matter action, a cosmological constant, and  a zeroth order 4-volume term) and the part $S_{ms}$ that essentially contains the background volume element $\tilde{\eta}$ and also has a  zeroth order 4-volume term.  As it turns out, the mass term is built out of \emph{both} of the algebraic part of $S_{0} $ (the cosmological constant and 4-volume term) and the purely algebraic term $S_{ms}.$   In comparison to derivations using the canonical stress-energy tensor \cite{FreundNambu,FMS}, the metric stress-energy tensor is much cleaner, in some respects more illuminating, but in some respects less transparent. The canonical tensor derivation is noticeably simpler in the known $w=\frac{1}{2}$ case for Freund and Nambu than in the other cases, for reasons that will be explained below.

Again we postulate universal coupling in the form
\begin{eqnarray}
\frac{\delta S}{\delta \tilde{\gamma} } = \frac{\delta S_{f} }{\delta \tilde{\gamma} } +k
 \frac{\delta S}{\delta \tilde{\eta} }.  
\end{eqnarray}  
Changing to the bimetric variables $\tilde{g}$ and $\tilde{\eta}$ implies, as before, that
\begin{eqnarray}
0 =  \frac{\delta S_{f}}{\delta \tilde{\gamma} }   +k \frac{\delta S}{\delta \tilde{\eta}} | \tilde{g}.
  \end{eqnarray}
Now we  introduce the relations $S_{f}= S_{f0} + S_{fm}$ and $S=S_{0} + S_{ms}$ to treat separately the pieces that existed in the massless case from the innovations of the massive case.  Thus 
\begin{equation}
\frac{\delta S_{f0}}{\delta \tilde{\gamma} } + 
\frac{\delta S_{fm}}{\delta \tilde{\gamma} }  =  -k \frac{\delta S_{0} }{\delta \tilde{\eta}} | \tilde{g} -k \frac{\delta S_{ms} }{\delta \tilde{\eta}} | \tilde{g}.
\end{equation}
Given the assumption that the new terms $S_{fm}$ and $S_{ms}$ correspond, this equation separates into the familiar part $\frac{\delta S_{f0}}{\delta \tilde{\gamma}}  =  -k \frac{\delta S_{0} }{\delta \tilde{\eta}} | \tilde{g} $ as before and the new part 
$$ \frac{\delta S_{fm}}{\delta \tilde{\gamma} } =  -k \frac{\delta S_{ms} }{\delta \tilde{\eta}} | \tilde{g}.$$
 $S_{0} $   is given by $S_{0} = S_{1} [\tilde{g}_{\mu\nu}, u] +   S_{2}$
as in the massless case.  Once again, we choose the simplest case and get the Hilbert action along with a cosmological constant, with matter coupled only to the curved metric, along with various terms that do not affect the equations of motion.
Assuming the free field mass term to be quadratic in the gravitational potential,
\begin{equation} \mathcal{L}_{fm} =-\frac{m^2}{2} \tilde{\gamma}^2  \sqrt{-\eta}^{1-2w},
\end{equation}
 its contribution to the field equation is
$$ \frac{\delta S_{fm}}{\delta \tilde{\gamma}  }  = - m^2 \sqrt{-\eta}^{1-2w} \tilde{\gamma}.$$ Changing to the bimetric variables gives
\begin{equation} \frac{m^2}{k} \tilde{g} \tilde{\eta}^{\frac{1}{w} -2} - \frac{m^2}{k}  \tilde{\eta}^{\frac{1}{w} -1}
 = k \frac{\delta S_{ms} }{\delta \tilde{\eta}} | \tilde{g}.
\end{equation} 

\subsection{Cases with $w \neq 1,$ $w \neq 0$}

The goal for  $S_{ms}$ is to obtain a mass term, so one can omit $u$ and $\hat{\eta}_{\mu\nu}$ from the `constant' of integration, leaving a function of $\sqrt{-g}$ only---which must be linear in order to be a scalar density of weight $1.$
Thus $$S_{ms} = \int d^{4}x \left( A \tilde{g}^\frac{1}{w} +  \frac{w m^2}{k^2 (1-w)} \tilde{g} \tilde{\eta}^{\frac{1}{w} -1} - \frac{w m^2}{k^2}  \tilde{\eta}^{\frac{1}{w}} \right) $$
as long as $w \neq 1.$  (The case $w=1$ yields a theory as well, but must be treated separately.)
Requiring $S_{ms}$ to vanish to zeroth order in $\tilde{\gamma}$ yields $$A = \frac{ w^2 m^2}{(w-1) k^2}.$$
Requiring $S_{ms}$ to vanish to first order, which is important for the field equations, gives nothing new.
For the second and higher order terms, the binomial series expansion 
yields
$$\mathcal{L}_{ms} = m^2 \sqrt{-\eta} \left(  - \frac{ \tilde{ \gamma}^2 }{2 \tilde{\eta}^2 }  
- \frac{[1-2w]k \tilde{\gamma}^3}{6 w \tilde{\eta}^3 }
+ \ldots \right).$$
For the free-field limit $k \rightarrow 0$ this expression reduces to the expected quadratic expression.  For the special case $w = \frac{1}{2},$ which turns out to be the Freund-Nambu theory (the only case previously obtained), the mass term contains no interaction part; the free mass term, quadratic in a weight $\frac{1}{2}$ potential, does not depend on the volume element in order to make a covariant (that is, weight $1$) Lagrangian density and so contributes nothing to the metric stress energy tensor.  (This perturbative expansion indicates nothing odd about the $w=1$ case, though the above integration was not permissible in that case.)
In terms of bimetric variables, the mass term for $w\neq 1$ is 
$$\mathcal{L}_{ms} = \frac{ w^2 m^2}{(w-1) k^2} \tilde{g}^\frac{1}{w} +  \frac{w m^2}{k^2 (1-w)} \tilde{g} \tilde{\eta}^{\frac{1}{w} -1} - \frac{w m^2}{k^2}  \tilde{\eta}^{\frac{1}{w}}.
$$
The factor $\tilde{g}^\frac{1}{w}$ often can be treated using a binomial series expansion; the series converges for $|k \tilde{\gamma} / \tilde{\eta} | <1$ \cite{Jeffrey}.  (A strong-field expansion is also possible, but will not be employed here.)
One can show that the binomial series expansion for the theory labeled by $w$ in terms of the weight $w$ variables is
\begin{equation}  \mathcal{L}_{ms} = - \frac{m^2 \sqrt{-\eta} }{k^2}  \sum_{j=2}^{\infty} 
 \left(k \frac{ \tilde{\gamma} }{ \tilde{\eta}  } \right)^j        \frac{ (\frac{1}{w} -2)!  }{ (\frac{1}{w} -j)! j!  }.
\end{equation} 
Here the expression $$\frac{ (\frac{1}{w} -2)!  }{ (\frac{1}{w} -j)! }$$ is shorthand for $(\frac{1}{w} -2) (\frac{1}{w} -3) \cdots (\frac{1}{w} -j+1);$ one need not make sense of the numerator and denominator separately in terms of Gamma functions, though one could do so.  This form is clearly well behaved in the vicinity of $w=1,$ so one can find the limit as $w \rightarrow 1$ to be
\begin{equation}  \mathcal{L}_{ms,w=1} = - \frac{m^2 \sqrt{-\eta} }{k^2}  \sum_{j=2}^{\infty} 
 \left(-k \frac{ \tilde{\gamma} }{ \tilde{\eta}  } \right)^j        \frac{ 1}{ j(j-1) }.
\end{equation} 
Note that this series expression of the theories leaves them not readily commensurable, due to the use of a different potential, bearing different relations to the more physically meaningful $\sqrt{-g}$ and $\sqrt{-\eta}$, for each value of $w$.  The use a $w$-specific potential in the derivation of each theory is quite important in the context discovery, however.  

\subsection{Case $w=1$}

The case  $w=1$ can be considered now.  The equation to integrate is 
\begin{equation}  \frac{m^2}{k} \tilde{g} \tilde{\eta}^{-1} - \frac{m^2}{k}  
 = k \frac{\delta S_{ms} }{\delta \tilde{\eta}} | \tilde{g}.
\end{equation} 
Performing the integration introduces a logarithm:
$$\mathcal{L}_{ms} = \frac{  m^2}{ k^2} \tilde{g}[ln(\tilde{\eta}) + f(\tilde{g})]   - \frac{ m^2}{k^2}  \tilde{\eta}.  $$  
In the interest of getting a scalar density, one can set the `constant' of integration $f(\tilde{g})$ to be
$$f(\tilde{g}) = a - ln(\tilde{g}) $$
for some constant $a.$ For $ w \neq 1$ the quadratic and higher terms came from the formal cosmological constant term proportional to $\sqrt{-g}$, but in this case that term is linear in the gravitational potential (in this choice of field variables), and hence serves  to cancel the noxious linear part of the mass-yielding nonlinear expression $ -\frac{  m^2}{ k^2} \tilde{g}  ln \left(\frac{ \tilde{g}}{ \tilde{\eta} } \right).$ This cancellation is a second service formed by the choice of $-1$ for the coefficient in $f.$  (This remark should be taken as merely heuristic, because use of nonlinear field redefinitions, such as re-expressing this mass term using a potential of a different density weight, or using the $w=0$ exponential field redefinition, would lead to a different sort of bookkeeping.)  Requiring the zeroth order part of the mass term to vanish as well gives $a=1.$  Using the Taylor expansion  $ln(1+x) = x - \frac{x^2}{2} + \frac{x^3}{3} -\ldots$, convergent for $-1<x\leq 1$  \cite[p. 564]{Shenk}, 
one obtains
$$\mathcal{L}_{ms} = m^2  \left(  - \frac{ \tilde{ \gamma}^2 }{2 \tilde{\eta} }  
+ \frac{k \tilde{\gamma}^3}{6  \tilde{\eta}^2 }
+ \ldots \right),$$
matching the expansion above to this order.  The full series expansion (for any $w,$ with possible exception of $w=0$---but that case will be vindicated shortly) can be shown to be 
\begin{equation}  \mathcal{L}_{ms,w=1} = - \frac{m^2 \sqrt{-\eta} }{k^2}  \sum_{j=2}^{\infty} 
 \left(-k \frac{ \tilde{\gamma} }{ \tilde{\eta}  } \right)^j        \frac{ 1}{ j(j-1) },
\end{equation}
in agreement with the expression above for the limit of the $w \neq 1$ family in the limit as $w \rightarrow 1.$ Thus the family of massive universally coupled scalar gravities is indeed continuous across $w=1,$ despite the need for special treatment of this case.  

One can also treat the case $w=1$ using  l'H\^{o}pital's rule for the indeterminate form $\frac{0}{0}.$  One has
\begin{eqnarray}
\lim_{w \rightarrow 1} \frac{m^2}{64 \pi G} \left[ \frac{ \sqrt{-g} }{w-1}  + \frac{ \sqrt{-g}^w \sqrt{-\eta}^{1-w} }{ w(1-w)} - \frac{ \sqrt{-\eta} }{ w}    \right]  \nonumber \\ =
\lim_{w \rightarrow 1} \frac{m^2}{64 \pi G} \left[ \frac{ w\sqrt{-g} - \sqrt{-g}^w \sqrt{-\eta}^{1-w}  - (w-1)  \sqrt{-\eta}   }{w^2-w}  \right] = \nonumber \\ 
\lim_{w \rightarrow 1} \frac{m^2}{64 \pi G} \left[ \frac{\sqrt{-g} - \sqrt{-g}^w  \sqrt{-\eta}^{1-w} (ln\sqrt{-g}  -   ln\sqrt{-\eta})  -   \sqrt{-\eta}   }{2w-1}  \right] \nonumber \\ =
 \frac{m^2}{64 \pi G} \left[ {\sqrt{-g} - \sqrt{-g}  ln\sqrt{-g}  +  \sqrt{-g}ln\sqrt{-\eta}  -   \sqrt{-\eta}   } \right]
\end{eqnarray}  
where the formula for exponentials of a non-natural base introduces the logarithms.  This expression of course agrees with that given above.

\subsection{Case $w=0$}

The case  $w=0$ is much more problematic, given the above bimetric field redefinition and the expression of universal coupling in terms of the derivative with respect to a volume element of some weight.  The weight 0 power of volume element is just 1, hardly a good field with respect to take a functional derivative.  The additive field redefinition defining $\tilde{g}$  appears to fail also.  From the Newtonian limit it follows that
$$ k^2 = 64 \pi G w^2.$$
To assess continuity of the field redefinition, one needs to know what happens to the meaning of $\tilde{\gamma}$ as $w \rightarrow 0.$  By considering 
$$ 1= \tilde{g}_w \tilde{g}_{-w} = (\tilde{\eta}_w + k_w \tilde{\gamma}_w) (\tilde{\eta}_{-w} + k_{-w} \tilde{\gamma}_{-w}) \approx 1  + \tilde{\eta}_w k_{-w} \tilde{\gamma}_{-w} + k_w \tilde{\gamma}_w  \tilde{\eta}_{-w} $$ near $w=0,$ one sees that the physical significance of the potential $\tilde{\gamma}$ does not jump discontinuously at  $w=0$ as long as  $k$ and $w$ have the same sign, which I choose to be positive for positive density weights $w.$  Thus  $k = 8 \sqrt{\pi G} w.$  But with $k$ proportional to $w,$ it appears that the bimetric field redefinition 
$$\tilde{g}_w = \tilde{\eta} + k(w) \tilde{\gamma}_w$$
(where the dependence on $w$ has now been made explicit) 
reduces to $1=1$ for $w=0.$  The universal coupling postulate 
\begin{eqnarray}
\frac{\delta S}{\delta \tilde{\gamma} } = \frac{\delta S_{f} }{\delta \tilde{\gamma} } +k
 \frac{\delta S}{\delta \tilde{\eta} }  
\end{eqnarray}  
suffers not only from the meaninglessness of $\frac{\delta S}{\delta \tilde{\eta} },$ but also from the vanishing of the coupling constant due to the linearity of $k$ in $w.$  

While these problems seem rather disastrous, in fact they are all soluble.  First we recall the series expansion above, with $k$ now expressed in terms of $w:$
\begin{equation}  \mathcal{L}_{ms} = - m^2 \sqrt{-\eta}   \sum_{j=2}^{\infty} 
( 8 \sqrt{\pi G})^{j-2}  \left( \frac{ \tilde{\gamma} }{ \tilde{\eta}  } \right)^j        \frac{ (1-2w)(1-3w) \cdots (1-jw +w)}{ j!  },
\end{equation} 
which has well behaved and simple coefficients as $w \rightarrow 0$. 
It is natural to drop the tilde on $\gamma$ and set $\tilde{\eta}$ to 1 for $w=0,$ leaving the simple form
\begin{equation}
\mathcal{L}_{ms,w=0} = - m^2 \sqrt{-\eta}   \sum_{j=2}^{\infty} 
( 8 \sqrt{\pi G})^{j-2}    \frac{  \gamma^j }{ j!  }.
\end{equation} 
This is clearly the sum of the quadratic and higher terms for the exponential function, so one infers
\begin{equation}
\mathcal{L}_{ms,w=0} = - \frac{ m^2 \sqrt{-\eta} } {64 \pi G } \left[ -1 - 8 \sqrt{\pi G} \gamma + 
exp   ( 8 \sqrt{\pi G} \gamma)  \right].
\end{equation} 
It remains to find a meaningful and appropriate notion of universal coupling that permits, one hopes, the derivation of an expression equivalent to this $w=0$ series.

The form of this series for $w=0$ suggests that one might think in terms of exponentials or logarithms to find a suitable field redefinition.  While the linear redefinition 
$$\tilde{g} = \tilde{\eta} + 8 \sqrt{\pi G} w \tilde{\gamma}$$
fails for $w=0,$ the $\frac{1}{w}$th root 
\begin{eqnarray} \sqrt{-g} = \tilde{g}^\frac{1}{w} = (\tilde{\eta} + 8 \sqrt{\pi G} w \tilde{\gamma})^\frac{1}{w} =
\sqrt{-\eta} \left(1 + 8w \sqrt{\pi G} \frac{ \tilde{\gamma} }{\tilde{\eta} } \right)^\frac{1}{w}  
\end{eqnarray}
makes sense for $w=0$ also.  The limit is
$$ \sqrt{-g} =\sqrt{-\eta} exp(8 \sqrt{\pi G} \gamma) .$$
An exponential change of variables very much like this was already employed by Kraichnan, though without application to massive theories \cite{Kraichnan}.  

By writing the trace of the stress-energy tensor in two different ways, one can show that the problems of meaningless field variable and of vanishing coupling also can be resolved.  The flat metric $\eta_{\mu\nu}$ can be written as
$$ \eta_{\mu\nu} = \hat{\eta}_{\mu\nu} \tilde{\eta}^\frac{1}{2w}.$$ 
Thus one recalls  that $$ \frac{\delta S}{\delta \tilde{\eta} } = \frac{1}{2w \tilde{\eta} } \frac{ \delta S}{\delta \eta_{\mu\nu} } \eta_{\mu\nu}.$$
Using this result in the postulate of universal coupling, the dependence on $w$ cancels out, giving for $w=0$
\begin{eqnarray}
\frac{\delta S}{\delta \gamma } = \frac{\delta S_{f} }{\delta \gamma } + 8 w \sqrt{\pi G}
\frac{1}{2w   } \frac{ \delta S}{\delta \eta_{\mu\nu} } \eta_{\mu\nu} =
\frac{\delta S_{f} }{\delta \gamma } + 4 \sqrt{\pi G}
 \frac{ \delta S}{\delta \eta_{\mu\nu} } \eta_{\mu\nu},
\end{eqnarray}
which makes sense even for $w=0.$ 
It is convenient to choose the weight $1$ variable  $\sqrt{-\eta},$ in terms of which universal coupling is $$\frac{\delta S}{\delta \gamma } = 
 \frac{\delta S_{f} }{\delta \gamma } + \frac{8 \sqrt{\pi G} }{\sqrt{-\eta} } 
 \frac{ \delta S}{\delta \sqrt{-\eta } }.
$$
The exponential change of variables, while leaving $\hat{\eta}_{\mu\nu} $ and $u$ alone, gives
\begin{eqnarray}
\frac{\delta S}{\delta \sqrt{ -\eta} } |\gamma  = \frac{\delta S}{\delta \sqrt{ -g} }  
\frac{ \sqrt{-g} }{ \sqrt{ -\eta} }  + \frac{\delta S}{\delta \sqrt{ -\eta} } |g, \nonumber  \\
\frac{\delta S}{\delta \gamma }  =  8 \sqrt{\pi G} \frac{\delta S}{\delta \sqrt{ -g} }  \sqrt{ -g}.
\end{eqnarray}
Installing these results in the universal coupling postulate yields
\begin{equation}
0 = \frac{ \delta S_f }{ \delta \gamma} +  8  \sqrt{\pi G} \sqrt{-\eta} \frac{\delta S}{\delta \sqrt{ -\eta} } |g,
\end{equation}
a result that is surprisingly indifferent to the non-additive form of the field redefinition. 
Letting the action $S$ be a sum of $S_1 + S_2$ from the massless case and $S_{ms}$ for the mass term,
one has
\begin{equation}
0 =  -m^2 \sqrt{-\eta} \gamma +  8  \sqrt{\pi G} \sqrt{-\eta} \frac{\delta S_{ms}}{\delta \sqrt{ -\eta} } |g.
\end{equation}
Making the change of variables in $S_f$ as well yields
\begin{equation}
0 =  - \frac{ m^2 \sqrt{-\eta} }{ 8 \sqrt{\pi G } } ln\left(\frac{\sqrt{-g} }{\sqrt{-\eta} } \right) +  8 \sqrt{\pi G} \sqrt{-\eta} \frac{\delta S_{ms}}{\delta \sqrt{ -\eta} } |g.
\end{equation}
Dividing by $\sqrt{-\eta}$ and integrating gives
$$ -64 \pi G \mathcal{L}_{ms} = m^2 \left[ -\sqrt{-\eta} ln\sqrt{-g} + \sqrt{-\eta} ln\sqrt{-\eta} -\sqrt{-\eta} +f(g) \right],
$$
where $f(g)$ is a `constant' of integration.  To get a scalar action, the obvious choice is $b \sqrt{-g}$ for some constant $b$.  Requiring the action to vanish to zeroth order yields $b=1;$  it vanishes to first order as well.  The result is 
\begin{eqnarray}  \mathcal{L}_{ms} = - \frac{m^2}{64 \pi G } \left[ - \sqrt{-\eta} ln\left( \frac{ \sqrt{-g} }{\sqrt{-\eta} }\right) + \sqrt{-g} -\sqrt{-\eta}  \right]= \nonumber \\ - \frac{ m^2 \sqrt{-\eta} } {64 \pi G } \left[     - 8 \sqrt{\pi G} \gamma +   exp   ( 8 \sqrt{\pi G} \gamma) -1 \right],
\end{eqnarray}
which was already obtained above as the $w \rightarrow 0$ limit of the series derived for the $w \neq 0.$  Thus the $w=0$ case in fact makes perfectly good sense and yields the theory that the $w \rightarrow 0$ limit leads one to expect. 
If Kraichnan had considered massive scalar gravity, then he would have obtained the $w=0$ theory readily.

One can also treat the case $w=0$ using  l'H\^{o}pital's rule for the indeterminate form $\frac{0}{0}.$  One has
\begin{eqnarray}
\lim_{w \rightarrow 0} \frac{m^2}{64 \pi G} \left[ \frac{ \sqrt{-g} }{w-1}  + \frac{ \sqrt{-g}^w \sqrt{-\eta}^{1-w} }{ w(1-w)} - \frac{ \sqrt{-\eta} }{ w}    \right]=  \nonumber \\
\lim_{w \rightarrow 0} \frac{m^2}{64 \pi G} \left[ \frac{ w\sqrt{-g} - \sqrt{-g}^w \sqrt{-\eta}^{1-w}  - (w-1)  \sqrt{-\eta}   }{w^2-w}  \right] = \nonumber \\
\lim_{w \rightarrow 0} \frac{m^2}{64 \pi G} \left[ \frac{\sqrt{-g} - \sqrt{-g}^w   \sqrt{-\eta}^{1-w}(ln\sqrt{-g} - ln\sqrt{-\eta})  -   \sqrt{-\eta}   }{2w-1}  \right] = \nonumber \\ 
  \frac{m^2}{64 \pi G} (   -\sqrt{-g} + \sqrt{-\eta} ln\sqrt{-g}   -  \sqrt{-\eta}ln\sqrt{-\eta}  + \sqrt{-\eta}  ), 
\end{eqnarray}  
in agreement with the formula given above.

 To sum up, while the $w=1$ case needed some special treatment and the $w=0$ case needed a great deal of special treatment, every real value of $w$ yields a (distinct) universally coupled massive scalar gravity.  
Thus we have found uncountably infinitely many massive scalar gravities, all derived by universal coupling, that give finite-range rivals to Nordstr\"{o}m's massless scalar theory.  These theories all provide a relativistic embodiment of the Seeliger-Neumann finite-range modification of Newtonian gravity.  There might be still other massive scalar gravities worthy of discovery, so no claim of exhaustiveness is made.

\subsection{All Cases in $w=0$ Exponential Variables}

Above the infinite family of theories was presented both using the bimetric variables $\sqrt{-g}$ and $\sqrt{-\eta}$ and using a series expansion of each theory in its own adapted perturbative field $\tilde{\gamma}_w.$  Recently it was found that the $w=0$ case suggests the relationship 
$$ \sqrt{-g} = exp(8 \sqrt{\pi G} \gamma) \sqrt{-\eta};$$
this field $\gamma$ (with no $\tilde{}$ and no density weight) is a neutral, ecumenical choice for expressing all of the massive gravities in a commensurable fashion---as compared to the series expansions above, which use different fields for different theories.  The result is
$$\mathcal{L}_{ms} = \frac{ m^2 \sqrt{-\eta} }{64 \pi G} \frac{ [ w e^{8 \gamma \sqrt{\pi G}} - e^{8w \gamma \sqrt{\pi G}} + 1-w ]}{w(w-1)}$$
for $w \neq 0, 1;$  these special cases are readily handled by l'H\^{o}pital's rule.
Using the series expansion for the exponential function, which converges everywhere, one has
$$\mathcal{L}_{ms,w} = -\frac{ m^2 \sqrt{-\eta} }{64 \pi G}   \sum_{j=2}^{\infty}  \frac{ (8 \gamma \sqrt{\pi G})^j}{j!} \sum_{i=0}^{j-2} w^i.$$

I will not attempt to \emph{derive} all the infinitely many theories using the $w=0$ exponential change of variables.  While such a derivation must be possible in some sense, the 
premises might look contrived by virtue of the apparently non-linear form of some of the terms.   Thus the role of the $w$-adapted field variables in the context of discovery is evident.  They allow infinitely many derivations to succeed using a manifestly free field \emph{via} a quadratic Lagrangian density, coupled to the total stress-energy tensor's trace, without powers of $\gamma$ in the coefficients.


%
%
%
%
%

\section{Stability}

In the interest of avoiding runaway solutions due to a potential energy with no lower bound, one wants 
to investigate the behavior of the algebraic mass/self-interaction term.  One might worry, for example, about theories  such that this  potential behaves like an odd polynomial (or worse) for large values (positive,  or negative unless the singularity as $\sqrt{-g} \rightarrow 0$ matters--in fact it will prove helpful) of the gravitational field $\gamma$ or some relative thereof \cite{VenezianoScalarBoson}.   While such strong values might invalidate the assumed validity of perturbation expansions sometimes made in this paper, a perturbative treatment at least suggests where trouble-spots might be found.  For theories with the self-interaction potential behaving like an even polynomial (or certain kinds of infinite series),  that sort of instability is not an issue, but  correct physical interpretation requires checking whether $\gamma = 0$ or the like is the true vacuum \cite{VenezianoScalarBoson}.   Checking these issues for all values of $w$ would be a substantial task, but it is not difficult to check some interesting cases.   Veneziano remarks that the Freund-Nambu theory is satisfactory on this count; I observe that for sufficiently negative values of the field  there is a crushing singularity, but the mass term repels from it.  

 While it is possible to check various isolated cases, treating all the infinitely many theories in a perturbative manner is not viable.  There is the further drawback, which takes a disjunctive form, depending on the choice of field variable.  In the $w$-adapted variables $\tilde{\gamma}$, the gravitational field  means different things for different theories.  On the other hand, if one uses the neutral $w=0$ field, then all theories give an infinite series---no case is a polynomial, making the analysis difficult.  

Fortunately one can avoid perturbative treatments altogether and extremize the mass-interaction part of the Lagrangian with respect to $\sqrt{-g}.$  One readily finds that the only critical point is the expected vacuum $\sqrt{-g}=\sqrt{-\eta}$ (which gives $\tilde{\gamma}=0$ for all $w$) and that it is indeed the ground state for all $w.$  In this sense massive scalar gravity is stable for all values of $w$.  Some of the theories repel infinitely from the singularity $\sqrt{-g}=0,$ while others do not.


%

\section{Why Canonical Tensor Derivation  Is Simple Only for $w=\frac{1}{2}$ Theory}

It is not difficult to see why Freund and Nambu discovered in effect the $w=\frac{1}{2}$ theory but not any of the other massive scalar gravities found here.  They use the canonical stress-energy tensor in its standard simple form, as in their equation 3b, where the trace is given as
$$ \frac{ \partial \mathcal{L} }{ \partial \phi,_{\mu}  } \phi,_{\mu} -4\mathcal{L}.$$  It is well known that   
one is permitted to add terms with automatically vanishing divergence to the stress-energy tensor, terms sometimes called ``curls'' \cite{Anderson} by virtue of their resemblance to the vector calculus theorem that the divergence of a curl is $0$ (itself a consequence of $1-1=0$).  It is quite understandable that in a 3-page paper Freund and Nambu did not consider this option (though terms something like this were studied, still  in a non-gravitational context, by Mack, by Chang and Freund,  and by Aurilia \cite{MackDilatation,ChangFreundScalar,AuriliaBrokenConformal}).    The derivation of the above massive scalar gravities using the canonical tensor for arbitrary $w,$ one can show, requires in the trace of the canonical stress-energy tensor the d'Alembertian of the term 
$$ \frac{ \tilde{\gamma} }{2 \sqrt{\pi G}} -     \frac{ (1 + 8w \sqrt{\pi G} \tilde{\gamma})^{\frac{1}{2w} }}{ 8 \pi G }  + \frac{1}{8 \pi G}, $$
as expressed in Cartesian coordinates, the use of which id advantageous  when the canonical stress-energy tensor is employed.  This extra term, which has second derivatives, vanishes if and only if $w=\frac{1}{2}.$  Thus the neglect of this term will cause the derivation to fail except in the case $w=\frac{1}{2}.$

The scalar gravity theory of Dehnen and Frommert \cite{DehnenMassiveScalar} is equivalent to the Freund-Nambu theory \cite{FreundNambu}  if one restricts the latter to matter fields  with conformally invariant kinetic terms.  This class includes not only electromagnetism and Yang-Mills theories (spin $1$), but also fermions (spin $\frac{1}{2}$).  It does not include standard scalar fields, however, though they do not remark on that important limitation.  With $\phi$ being gravity and $\chi$ being matter, Freund and Nambu find that a standard scalar field coupled to gravity has an interaction involving 
$\phi (\partial \chi)^2.$  Dehnen and Frommert assume without justification that no such terms exist, as well as assuming that no nonlinear terms  $\phi (\partial \phi)^2$ arise.  The latter terms can be absorbed by a nonlinear field redefinition of the gravitational potential $\phi.$  For electromagnetism and Yang-Mills fields, taking the potentials to be covectors (as one usually does) removes the 
$\phi (\partial \chi)^2$  terms.  For spinors, I recall from above that redefining the spinor fields to have density weight $\frac{3}{8}$  removes the coupling of the gravitational potential $\phi$ to the spinor (in this case a term roughly like $\phi \psi \partial \psi$).  
The ability of the Dehnen-Frommert theory to accommodate spin $\frac{1}{2}$ and spin $1$ fields was exploited in some subsequent papers where only those matter fields were entertained \cite{DehnenHiggsScalar,DehnenMassiveScalarFermion}.   The Dehnen-Frommert derivation is not so simple because it involves multiple field redefinitions motivated  by the need to recover from the assumption of purely nonderivative coupling between gravity and matter or the desire to derive massive scalar gravity in a Higgs-looking fashion.

In view of the disadvantages of the (unsymmetrized, unimproved) canonical energy-momentum tensor for various fields---asymmetry, gauge dependence, and nonvanishing trace in some contexts where one might have wanted the trace to vanish \cite{ForgerRomerStress}---the use of the metrical definition is a good deal more convenient.  In principle one could use a Belinfante-Rosenfeld equivalence theorem   and employ the symmetric Belinfante tensor or the like.  However, it seems not very practical to derive an unknown  Lagrangian density for non-scalar matter fields by looking for solutions to a messy identity involving all sorts of partial derivatives of the Lagrangian density with respect to the field derivatives.  It seems not accidental that thus far only \emph{via} the metrical definition has an infinity, or even a variety, of universally coupled scalar gravities been derived.

As Josep Pons has recently recalled \cite{PonsEnergy}, the process for making a flat space-time theory formally generally covariant admits considerable freedom in choosing density weights for fields:  the comma-goes-to-semicolon rule thus has considerable ambiguity which tends to go unnoticed.  While for many purposes the choice of density weight does not matter much (other than affecting the forms of the Lie and covariant derivatives), the metrical stress-energy tensor, in particular its trace, is significantly affected.  Above I observed that  the use of a suitably densitized spinor allowed $\sqrt{-g}$ to disappear completely from the massless Dirac equation; I observe that  one can do the same thing using conformally invariantly coupled scalar fields.  The scalar field is replaced by a scalar density $\phi_w$ of weight 
$w= \frac{n-2}{2n}$ (which comes to $\frac{1}{4}$ in four space-time dimensions) by absorbing suitable powers of $\sqrt{-g}$; the pleasant result is that $\sqrt{-g}$ thereupon disappears completely from the theory.
The expression 
$$  \sqrt{-g}^\frac{2}{n} \left[\nabla^2 \phi_w - \frac{n-2}{4(n-1)} R \phi_w \right]$$
is the same for all conformally related metrics, from which it follows that $\sqrt{-g}$ simply cancels out altogether.  Multiplying this expression by $\phi_w$  gives a scalar density of weight $1,$ which is thus a suitable Lagrangian density.  One should be able to expand the expression out using $\hat{g}_{\mu\nu}$ and $\sqrt{-g}$ and watch the latter disappear; the  result will not \emph{look} like a scalar density, so the expression $  \sqrt{-g}^\frac{2}{n}[\nabla^2 \phi_w - \frac{n-2}{4(n-1)} R \phi_w]$ has some advantages.  One could also drop some total divergences to remove second derivatives of $\phi_w$ and/or the metric, if desired, perhaps at the expense of manifest covariance (somewhat like the Einstein $\Gamma\Gamma$ Lagrangian density for General Relativity).   The absence of $\sqrt{-g}$ immediately implies a traceless metrical stress-energy tensor even off-shell.  The use of densitized scalar and spinor fields thus allows one to identify and reject $\sqrt{-g}$ as surplus structure in some notable contexts.


%
%





\end{document}